\documentclass{aa}
\usepackage{pifont}
\usepackage{cancel}

\usepackage{graphicx}
\usepackage{bm}
\usepackage{soul,color}
\usepackage{comment}
\usepackage[varg]{txfonts}

\usepackage[hidelinks]{hyperref}

\renewcommand{\vec}[1]{\bm{#1}}

\newcommand{\tauline}{\ensuremath{\tau_\text{line}}}

\newcommand{\cmf}{\ensuremath{\leadsto}}
\newcommand{\nucmf}{\ensuremath{\nu^\cmf}}

\newcommand{\vinfty}{\ensuremath{V_\infty}}
\newcommand{\Rinfty}{\ensuremath{R_\infty}}

\defcitealias{fisak2023}{Paper~I}

\newcommand{\antares}{``Andy Antares''}

\newcommand{\modgrid}{model grid}
\renewcommand{\modgrid}{modGrid}
\newcommand{\propgrid}{propagation grid}
\renewcommand{\propgrid}{propGrid}

\newcommand{\jednad}{\mbox{1\protect\nobreakdash-D}}
\newcommand{\dvad}{\mbox{2\protect\nobreakdash-D}}
\newcommand{\trid}{\mbox{3\protect\nobreakdash-D}}
\newcommand{\CMFGEN}{{\tt CMFGEN}}
\newcommand{\FASTWIND}{{\tt FASTWIND}}
\newcommand{\PoWR}{{\tt PoWR}}
\newcommand{\Tardis}{{\tt Tardis}}
\newcommand{\Python}{{{\tt SIROCCO} (formerly Python)}}
\newcommand{\mpiamrvac}{{\tt MPI-AMRVAC}}

\newcommand{\vel}{\mathtt{v}}
\newcommand{\velb}{\vec{\mathtt{v}}}

\usepackage{tikz}
\usetikzlibrary{arrows,snakes,backgrounds}

\begin{document}

\title{Progress towards a 3D Monte Carlo radiative transfer code for
outflow wind modelling}
\subtitle{II. {\trid} applications}
 \author{J. Fi\v{s}\'{a}k
         \inst{1}
    \and
         J. Kub\'{a}t
         \inst{1}
    \and
        N. Moens
        \inst{2}
 	\and
	       B. Kub\'{a}tov\'{a}
 	     \inst{1}
    }
 \institute{
 Astronomical Institute of the Czech Academy of Sciences, Fri\v{c}ova 298,
 CZ-251 65, Ond\v{r}ejov, Czech Republic \\ \email{jakub.fisak@asu.cas.cz}\label{inst1}
\and
Instituut voor Sterrenkunde, KU Leuven, Celestijnenlaan 200D, 3001 Leuven, Belgium \label{inst2}
}


\abstract
{
The massive hot stars play crucial role in the dynamics of galaxies. These
stars influence their surroundings through strong winds which are highly
structured processes. The theoretical study of the non-symmetric phenomena of
the stellar winds are becoming important these days, mainly because {\jednad}
models are not sufficient enough.
}
{
We present a new version of our Monte Carlo radiative transfer code
{\antares} with improved treatment of the velocity field for arbitrary
geometries.
%
Our aim is to develop a numerical scheme that can incorporate a general velocity
field defined at discrete points.
%
Our main objective is to calculate
radiative transfer in a general input hydrodynamic model.
}
{
The \antares code
currently
calculates pure radiative transfer. The input model is
pre-calculated by another hydrodynamical code. The whole
radiative transfer
calculation is then
processed in a Cartesian grid.
%
%
%
Radiative transfer is solved using the Monte Carlo approach in {\trid}
regardless of the input hydrodynamical model's dimension.
The velocity field at any given point is interpolated using the trilinear
interpolation.
The optical depth is then integrated numerically along the photon's path.
}
{
We verified the accuracy of the numerical velocity interpolation by
comparison with results obtained for analytical velocity fields, achieving
successful outcomes.
We also tested the radiative transfer solution on a {\trid} model generated
from a {\dvad} hydrodynamic model, and obtained emergent radiation.
}
{
The code is suitable for the numerical solution of radiative transfer in {\trid}
with arbitrary velocity fields.
}
\maketitle

\section{Introduction}

Radiatively driven outflows from massive hot stars (O, B, and Wolf–Rayet stars)
are a cornerstone of stellar and galactic evolution. By removing angular
momentum and large amounts of mass, these winds reshape surface abundances,
alter evolutionary tracks, and regulate feedback into the interstellar
medium—thereby influencing everything from local \ion{H}{ii} regions to the
high-redshift ionizing budget. To interpret today’s high-quality UV, optical,
and X-ray observations, we need reliable radiative-transfer models of
massive-star winds. The mass loss from hot-star winds is crucial to
massive-star evolution, yet standard {\jednad} analyses can misestimate it
because real winds are highly structured -- densely clumped and porous in space
and velocity, and sometimes shaped by magnetic/rotational geometry -- which
modifies ionization and line formation; consequently, developing genuinely
{\trid} radiative-transfer (RT) models is now essential for robust diagnostics.

Over the last few years, theory has increasingly shown that {\jednad} hot-star
atmosphere/wind analyses are inadequate because rotation, magnetic confinement,
corotating interaction regions, and small-scale clumping/vorosity make the
outflows intrinsically {\trid} -- pushing us toward full {\trid} RT models for
trustworthy mass-loss diagnostics.  The \citet{moens2022a, moens2022b}
illustrates the origin of clumps in the atmospheres. The {\dvad} models were
studied by \citet{debnath2024}. Later, \citet{udDoulaEtal2025} and
\citet{narechaniaEtal2025} showed numerically that magnetic fields indeed also
can affect structure formations in these multi-D setups. Hot stars are studied
also in the ULLYSES project, such as \citet{sander2024}.

The standard {\jednad} spectral synthesis codes such as \FASTWIND, \CMFGEN,
\PoWR, {\dots} treat velocity field with several crucial assumptions. The main
approximation is an exclusion of non-monotonical velocity fields. Treatment of
general velocity field is significantly more complicated. Many codes, such as
{\Tardis} \citep{kerzendorf2014, Vogl2019, kerzendorf_wolfgang_2022}, the
{\Python} code (\citealt{matthewsEtal2025, long2002}, and
\citealt{higginbottom2013}), Monte Carlo clumping wind model \citep{surlan2012,
surlan2013}, the PoWR code \citep{grafener2002, hamann_grafener2003,
sanderEtAl2015} use analytical velocity fields. Analytical velocity fields are
simpler to treat than numerical ones because the latter require interpolation
procedures to be implemented. The non-analytical solutions are possible in the
{\Python} code mentioned earlier.

In this paper, we build on the work presented in our previous paper
\citep[hereafter \citetalias{fisak2023}]{fisak2023}, in which the Monte Carlo
code ``Andy Antares''\footnote{This code is available online \url{https://github.com/jfisak/andyAntares}} had been introduced.

We show some first applications on
the first full {\trid} models and we will describe important differences
between {\jednad} and multi-D cases. In Section \ref{Sec:methods} we describe
all used methods and calculations. Section \ref{Sec:testing} shows tests of
simple {\jednad} models solved using algorithms applicable to general {\trid}
problems. Finally, in Section \ref{Sec:applications} we present the very first
applications on full {\trid} ({\dvad}) models including calculation of angle
dependent spectra.

\section{Methods} \label{Sec:methods}

The code uses two basic grids, the model grid (hereafter \modgrid), and the
propagation grid (hereafter \propgrid). The {\modgrid} describes the underlying
hydrodynamic model atmosphere structure (e.g. temperature, density, velocity,
etc.) and may use any coordinates (cartesian, spherical polar, cylindrical,
etc.). The {\propgrid} is used for radiation transfer and uses exclusively the
cartesian coordinates. Relation between these two grids is described in detail
in \citetalias{fisak2023}. Hereafter, we define a `{\modgrid} cell' as a volume
in {\trid} space containing a point defined in the input file. In the {\jednad}
case, it can be a spherical shell, and in {\dvad} and {\trid} cases, it can be
a more complicated shape. The {\propgrid} cell is always a cube-shaped volume.

\subsection{Treatment of interaction in lines}
All considered atomic lines are stored in a single array. For each line, this
array contains atomic ion information (element index, ion index, lower and
upper level indexes), the line frequency, and the oscillator strength. A packet
will interact with the line(s) closest to the CMF frequency of the packet if
the condition (13) in \citetalias{fisak2023} is met -- their resonance
(Sobolev) points ($\nu^\cmf = \nu_\mathrm{line}$) are the closest to the
packet's current position. We assume that the velocity field changes smoothly
between model points, so the CMF frequency also changes smoothly.

The most effective way is to sort the line list in respect to the line
frequency. If we calculate the CMF frequency of a packet and if the frequency
is an increasing or decreasing function along the packet's path, we can find
the next (and the closest) line the packet will interact with. This is followed
by incrementing/decrementing of an index of the next line, instead of searching
in an unsorted array, which significantly saves a lot of computing time.

Another advantage of using the described method is a straightforward
application to non-monotonic velocity fields. Using one logical variable
detecting if the CMF frequency of a packet is increasing or decreasing, we can
simply select the correct line in the corresponding blue or red shift. This
possibility is an advantage of our code because other codes either  do not
include blue shifed parts properly or omit them completely. 

\subsection{Diagnostics of created {\propgrid}}
The created {\propgrid} is crucial for a proper description of the whole model.
Consequently, it must be tested, if the {\propgrid} correctly covers the
physical model (represented by the {\modgrid}) and if all inhomogeneities are
included in the calculation via the {\propgrid} as well. In this section, we
describe the methods used to test the created {\propgrid}.

Some tests have already been published in \citetalias{fisak2023} in Section 4.
Firstly we check, if all {\modgrid} cells are connected to at least one
{\propgrid} cell. This is called ``Relative covering'' in the paper
\citetalias{fisak2023} (see Table 3 in that paper). This number should be equal
to one.

The second method compares volumes of {\propgrid} cells and connected
{\modgrid} cells: the first equation (Eq.\,41 and 42 in \citetalias{fisak2023}),
\begin{equation}
 \tag{I.41}
 \chi_I = \frac{\displaystyle\sum_{\text{PC}\rightarrow I}V_\mathrm{PC}}{V_{\text{I}}},
\end{equation}
describes the ratio of the volumes of all {\propgrid} cells $\mathrm{PC}$
associated with {\modgrid} $I$ to the total volume of {\modgrid} I:
$V_{\text{I}}$. To make the results more clear, we calculated the standard
deviation of all {\modgrid} cells using Eq.\,(42) from \citetalias{fisak2023},
\begin{equation}
 \label{meanchidef}
 \tag{I.42}
 \sqrt{\left<\chi^2\right>} = \sqrt{\frac{1}{N_\text{\modgrid}}
 \sum_{I=1}^{N_\text{\modgrid}} \left(\chi_I - 1\right)^2},
\end{equation}
which should equal zero in an ideal case.

We used the relative covering test to check the created grid. The second method
is really tricky to calculate in {\dvad} and {\trid} cases, since we have to
calculate the Voronoi volumes for each point in multidimensional space. This
calculation will therefore be added in the future.

\subsection{Velocity field treatment}
\label{Ssec:velfield}

Most {\trid} hydrodynamic codes also provide full information about the
velocity structure, consequently, each grid point in {\modgrid} contains a velocity
vector. When solving radiative transfer in a moving environment, we 
need to obtain a velocity vector generally at any given point. This is not an easy
procedure, hence we have developed an algorithm based on the trilinear
interpolation method. 

A simple approximation of the velocity field via a step function (i.e. constant
velocity within a given {\propgrid} cell) is not acceptable. This can lead to
the incorrect treatment of line and continuum interactions since the continuum
optical depth depends on the path length over which interaction is possible.
The Doppler shift (hence the CMF frequency) has to be known exactly at each
point along the packet's trajectory. The position of the resonance points is
important for the determination of packet interactions with matter. If the
velocity field were described by an analytical function \citepalias[as, for
example, in][]{fisak2023}, the computation would be quite simple. However, the
general input velocity field can not be approximated by a simple analytical
function. Therefore, an interpolation technique must be implemented in the
code.

We used a method which obtains velocity vectors from connected {\modgrid} cells
during photon propagation. This method is described in detail in the
forthcoming paragraphs.

\subsubsection{Interpolation of the velocity field from the {\modgrid}}
\label{Ssub:intp_modgrid}

\begin{figure}
 \centering
 \includegraphics[width=.45\textwidth]{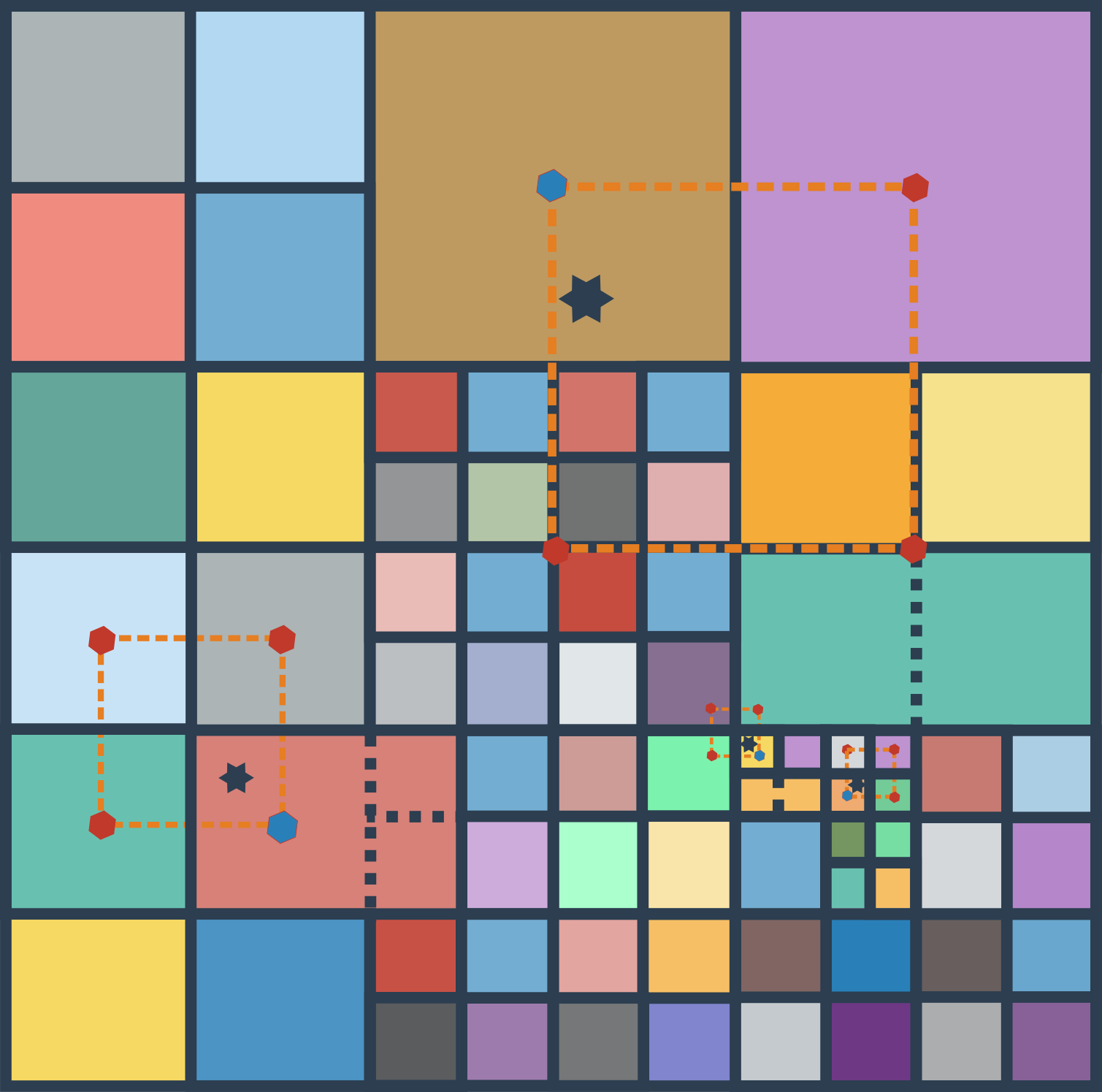}
 \includegraphics[width=.45\textwidth]{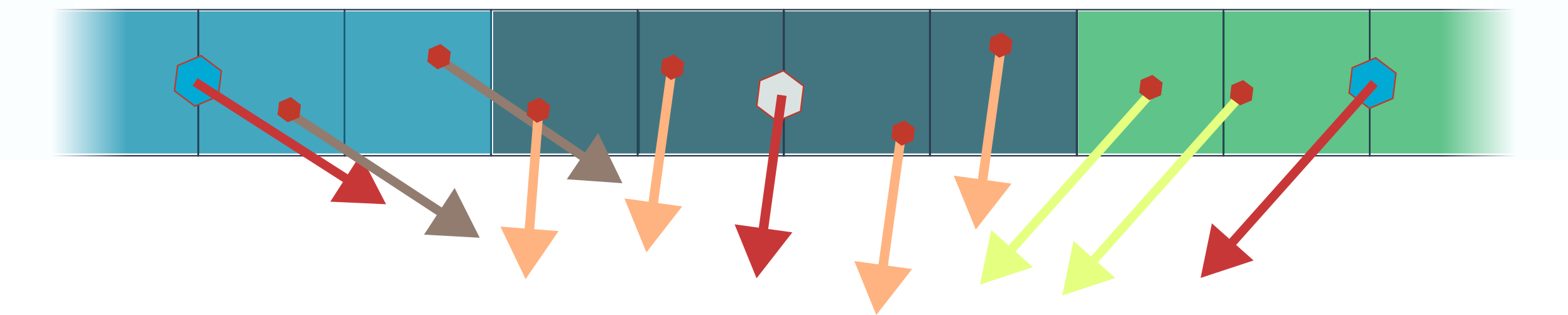}
 \caption{
  An illustrative example of an adaptive {\propgrid} with {\propgrid} cells
  belonging to different {\modgrid} cells. Each {\propgrid} cell belongs to any
  {\modgrid} cell, which is distinguished by different colours (the same colour
  of non-neigbouring cells does not mean the same associated {\modgrid} cell, we
  did not want to use many colours).
  \emph{Above}: An algorithm for choosing interpolation points based on the
  current {\propgrid} cell. The current {\propgrid} cell is the cell containing
  a dark star, which locates the place where we calculate the interpolated
  velocity vector. There are four such examples in the figure. The blue hexagons
  represent the central points of the current {\propgrid} cells (where the dark
  stars are located), the red hexagons represent the derived points used for the
  trilinear interpolation. If more than one {\propgrid} cell belongs to the
  same {\modgrid} cell, the boundary between them is drawn with a dotted line.
  \emph{Below}: The red arrows represent the velocity vectors defined in the input.
  The other arrows represent the velocity vectors used for trilinear interpolation
  at any particular point (in this picture represented by little red
  hexagons).}
 \label{Fig:interpvel}
\end{figure}

The method is illustrated at Fig.~\ref{Fig:interpvel}. It gets the velocity
vector directly from the {\modgrid} associated with the {\propgrid} cell in
question during packet propagation. First of all, the interpolation points (the
blue and red hexagons in the upper part of Fig.\,1) are obtained. These points
mark the corners of a cell of the same size as the current {\propgrid} cell,
with the blue hexagon corner at the current {\propgrid} cell centre. The
corners are positioned so that the dark star (the place where we calculate the
interpolated velocity vector) is inside the created object, which is marked by
an orange dashed line. The velocity vectors of the {\propgrid} cells are
defined as those of the corresponding {\modgrid} cells and are defined at each
corner of the orange cube (red hexagons). This velocity is the same as the
velocity of the {\modgrid} cell belonging to the given {\propgrid} cell
(distinguished by the different colours of the {\propgrid} cells). As can be
seen from Fig.~\ref{Fig:interpvel} below, if more {\propgrid} cells belong to
one {\modgrid} cell, the velocity vector is the same in all of these
{\propgrid} cells.\footnote{This only applies to the {\trid} model. For the
{\jednad} and {\dvad} models, the vector magnitudes are the same, but the
direction differs depending on the position, for example in the case of a
radial field.}
\subsection{Line interaction calculation}
The interaction in lines, or better said, the line optical depth calculation,
changes in the case of a more general velocity field. We cannot use the simple
equation \citepalias[Eq.\,10]{fisak2023} for these, because it is only valid
for spherically symmetric velocity fields. Instead, we have to use the more
general equation (still assuming the Sobolev approximation), which is of the
form
\begin{multline} 
 \tauline = n_{l} f_\mathrm{lu}\frac{\pi e^2}{m_\mathrm{e}c}
 \left(1 - \frac{n_u g_l}{n_l g_u}\right)\\
 \times\left(\frac{\text{d}s}{\text{d}\nucmf}\right)_{\nu_{lu}}
 \times 
 \begin{cases}
  0 & \nu_{lu}\notin [\nucmf(0), \nucmf(s_0)]\\
  -1 & \nu_{lu}\in [\nucmf(0), \nucmf(s_0)]
 \end{cases}
 \label{Eq:tauline}
\end{multline}
here (cf. Eq. (9) in \citetalias{fisak2023} or Eq.~4.15 in
\citealt{kromer2009}, where the Einstein coefficient $B_{lu}$ instead of the
oscillator strength $f_\mathrm{lu}$ is used). The indexes $l$ and $u$ denote
the lower and upper energy level, respectively; $n_l$ and $n_u$ denote the
respective level populations, $g_l$, $g_u$ are the statistical weights,
$B_{lu}$ is the Einstein B coefficient, $\nu_{lu}$ is the transition frequency
between these two levels. Finally, $s$ measures the distance from the current
position (geometrically, it is a line parameter), and $s_0$ denotes the
position of the resonance point.

From Eq.~\eqref{Eq:tauline} it is clear that the derivative
$\mathrm{d}s/\mathrm{d}\nucmf$ must be evaluated, which introduces several
additional numerical problems to solve.
\begin{figure}
 \centering
 \includegraphics[width=.45\textwidth]{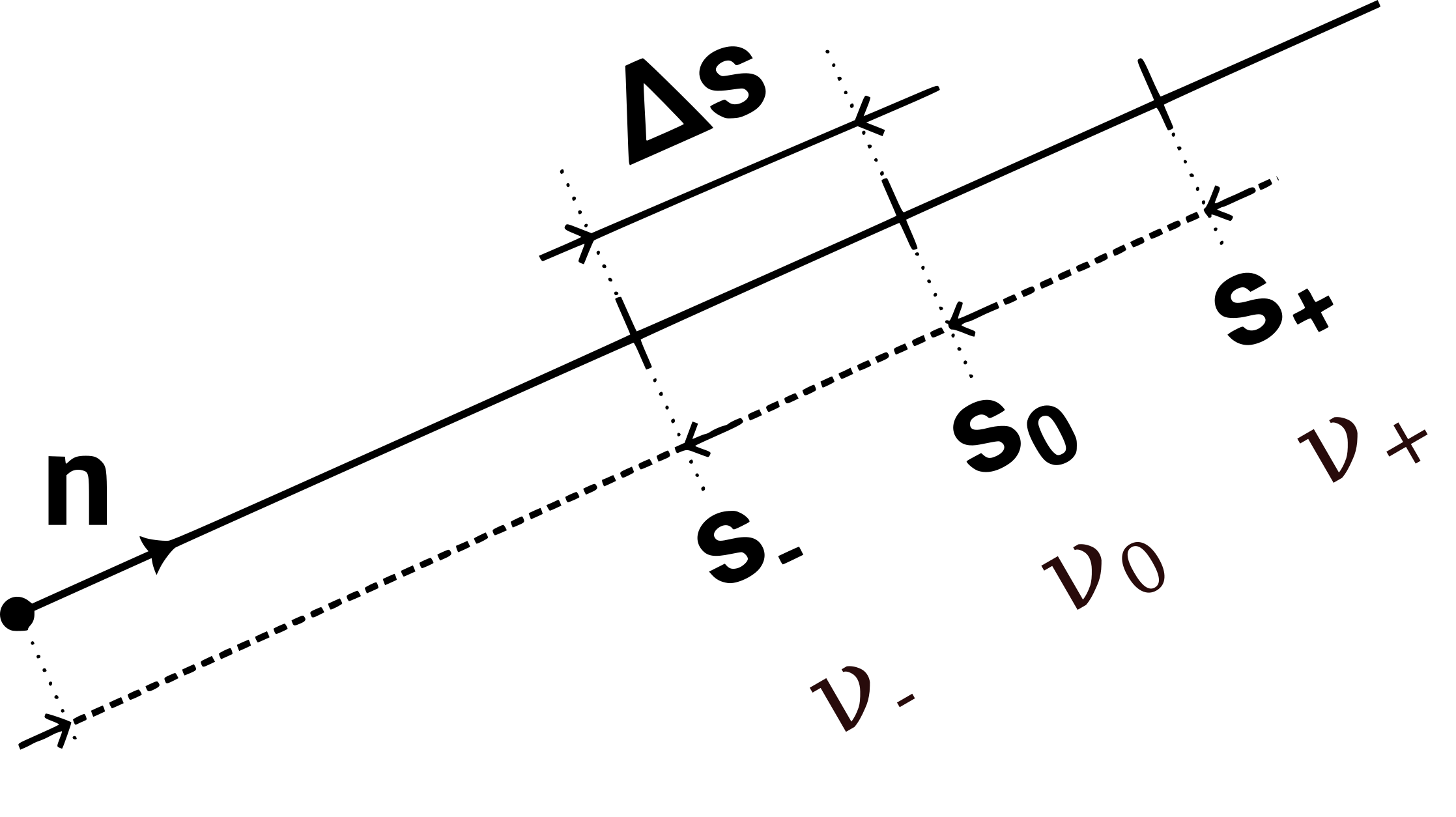}
 \caption{ The derivative $\mathrm{d}s/\mathrm{d}\nucmf$ is calculated using
  Eq.~\ref{Eq:tauline}. A packet is located in the beginning of the dashed line,
  the arrow points to its current direction. The resonance point is denoted by
  $s_0$ and the packet's CMF frequency is exactly equal to the line frequency.
  $s_-$ and $s_+$ denote positions of points with a distance $\Delta s$ from the
  resonance point.}
 \label{Fig:lineint}
\end{figure}
One of the problems is the choice of appropriate $\Delta s$, since at least two
points are needed to calculate a derivative. Its value at the point $s_0$ is
\begin{equation}
 \left(\frac{\text{d}s}{\text{d}\nucmf}\right)_{\nu_{lu}} =
 \frac{s_+-s_-}{\nucmf_+-\nucmf_-},
 \label{Eq:dsdnu}
\end{equation}
in the case, when $s_0-s_- = s_+-s_0 = \Delta s$ (the resonance point is
located exactly in the middle of points $s_+$ and $s_-$, see
Fig.~\ref{Fig:lineint}).

\section{Testing the generalized {\trid} calculations} \label{Sec:testing}

The MC calculations are processed in a {\trid} space, however, the equations
implemented in \citetalias{fisak2023} assume spherical symmetry of the models.
Here we generalise these equations for a {\trid} calculation. We test these new
{\trid} algorithms on spherically symmetric cases and compare the results with
calculations of the same models in {\jednad}.

In order to test the properties of {\trid} radiation transfer in more detail,
and to avoid being overwhelmed by the possible difficulties of a general
{\modgrid}, we use a regularly spaced {\modgrid}. This enables us to create the
{\propgrid} so that each of its cells contains one {\modgrid} cell.

We show simple tests of {\trid} procedures. The key factor is the velocity
interpolation, which is described in the next section. In the subsequent
section we test the calculation of the line optical depth which is more complex
in the case of {\trid} models.

\subsection{Velocity field}
\label{Sec:velfield}
In this part, we apply the method described in the Section \ref{Ssec:velfield},
and test the interpolation method \ref{Ssub:intp_modgrid}. We ran several
simple tests to evaluate the implementation of trilinear interpolation. The
first of these compares the analytical and numerically calculated velocity
magnitudes. For this purpose, we created a simple {\trid} model, its parameters
are described in Tab.~\ref{Tab:testing_hmodels}. In this model, each {\modgrid}
cell position is defined by a vector $\vec{r}$ with three independent
coordinates, $x$, $y$, and $z$, namely
\begin{equation}
 \vec{r} = \vec{r}_\mathrm{min} + 
  (i\cdot w_\mathrm{x},\, j\cdot w_\mathrm{y},\, k\cdot w_\mathrm{z}),
\end{equation}
where the indexes $(i, j, k)$ are integers from the set $\{1, 2, \cdots ,
N_d\}$, $d$ stands for $x$, $y$, or $z$. $N_d$ is defined by the input model
$N_x\times N_y\times N_z$, and
\begin{equation}
 \vec{r}_\mathrm{min} = (x_\mathrm{min},\, y_\mathrm{min},\, z_\mathrm{min})
\end{equation}
are the coordinates of the {\modgrid} corner ($x_\mathrm{min}$,
$y_\mathrm{min}$, and $z_\mathrm{min}$ are minimum values of the $x$, $y$, and
$z$ coordinates, respectively). The maximum coordinates of the {\modgrid} are
\begin{equation}
 \vec{r}_\mathrm{max} = \vec{r}_\mathrm{min} + 
  \left[(N_x-1)\cdot w_\mathrm{x},\, (N_y-1)\cdot w_\mathrm{y},\, (N_z-1)\cdot
  w_\mathrm{z}\right],
\end{equation}
where $\vec{w} = \left(w_\mathrm{x}, w_\mathrm{y}, w_\mathrm{z}\right)$ is the
distance between points in all three directions, simply calculated from the
total size of the {\modgrid} and the number of cells in a selected dimension,
$$w_d = \frac{(\mathbf{r_\mathrm{max}})_d-(\mathbf{r_\mathrm{min}})_d}{N_d -
1},$$ where $d$ stands for $x, y, z$.
\begin{table}
 \begin{tabular}{p{4cm} p{4cm}}
  \hline
    \textbf{parameter} & \textbf{value} \\
  \hline
   effective temperature $T_\mathrm{eff}$ & 14734 K\\
   stellar radius $R_*$ & 8072 $R_\odot$\\
   upper boundary $R_\infty$ & 5 $R_*$\\
   outer boundary velocity $V_\infty$ &
   $3\times 10^9\,\mathrm{cm\cdot s^{-1}}$\\
   $\beta$ parameter & 1.3 \\
  \hline
   ${N}_x\times {N}_y\times {N}_z$ (homologous)& $50\times 50\times 50$\\
   ${N}_x\times {N}_y\times {N}_z$ (beta-law)& $200\times 200\times 200$\\
   range of the {\propgrid} (same for $x$, $y$, and $z$) & 
    $(-5.1\cdot R_*, +5.1\cdot R_*)$ \\
   {\propgrid} cell width (same for $x$, $y$, and $z$) (homologous)&
   $\frac{1}{5}R_*$\\
   {\propgrid} cell width (same for $x$, $y$, and $z$) (beta-law)&
   $\frac{1}{20}R_*$\\
  \hline
   number of packets & 20000000 \\
  \hline
 \end{tabular}
 \caption{Parameters for the calculated models. The upper part shows the
  general setup of the computational domain, and the physical conditions at the
  boundaries. The middle part shows the properties of the regular {\propgrid},
  and the bottom part specifies the number of packets used in the calculation. }
 \label{Tab:testing_hmodels}
\end{table}
The velocity in each {\modgrid} cell
is calculated using the formula
\begin{equation}
 \velb(x_I, y_I, z_I) = \vel_{I} \frac{(x_I, y_I, z_I)}{\sqrt{x_I^2 + y_I^2 +
 z_I^2}},
 \label{Eq:hom_I}
\end{equation}
Here, $I$ is an index of a {\modgrid} cell $I\in\{1,2,\cdots , N_\mathrm{mg}\}$
(see \citetalias[Section 4]{fisak2023}), $v_I$ is the velocity value in the $I$-th
{\modgrid} cell, $N_\mathrm{mg} = N_x\cdot N_y\cdot N_z$, and finally, $x_I$,
$y_I$, and $z_I$ are Cartesian coordinates for the {\modgrid} cell with
index $I$. 

The {\propgrid} can be set very simply. The {\modgrid} is regular and we define
the {\propgrid} position so that each {\modgrid} cell is located at the
{\propgrid} cell centre. Therefore, each {\propgrid} cell corner can be defined
by a relation
\begin{equation}
 \vec{r}^\mathrm{PG} = \vec{r}^\mathrm{PG}_\mathrm{min} + 
  (i\cdot w_\mathrm{x},\, j\cdot w_\mathrm{y},\, k\cdot w_\mathrm{z}),
\end{equation}
where $\vec{r}^\mathrm{PG}_\mathrm{min} = \vec{r}_\mathrm{min} - \vec{w}/2$
(all variables defined by the {\modgrid}) and, as in the previous case,
$(i, j, k)$ are the integers $\{1, 2, \cdots , N_d\}$. In this case, 
\begin{equation}
 \vec{r}^\mathrm{PG}_\mathrm{max} = \vec{r}^\mathrm{PG}_\mathrm{min} + 
  \left[(N_x+1)\cdot w_\mathrm{x},\, (N_y+1)\cdot w_\mathrm{y},\, (N_z+1)\cdot
  w_\mathrm{z}\right],
\end{equation}
is the vector with the maximal values of coordinates. Note that {\propgrid} is
larger than {\modgrid} in this case.

For the testing purposes, we used two velocity fields. The homologous
approximation is a simple linear velocity law, whilst the beta law is a more
complicated yet still monotonic, function. The main advantage of these
functions is offering the possibility of direct comparison of the calculated
velocity with that obtained from an analytical function. We show the
differences in both cases.
\subsubsection{Homologous approximation}
In the homologous approximation, the velocity $\vel_I$ in Eq.~\eqref{Eq:hom_I}
is given by
\begin{equation}
 \vel_I = \sqrt{x_I^2+y_I^2+z_I^2} \frac{\vinfty}{\Rinfty}.
 \label{Eq:velhomol}
\end{equation}
The left panels of Fig.~\ref{Fig:vel_beta} show the analytical velocity in
comparison to the numerically calculated velocities (upper figure) and its
relative error (lower figure). The values at the majority of points fit the
analytical expression well, with the exception of points close to the lower
boundary.

\subsubsection{$\beta$-law}
\begin{figure*}
 \centering
 \includegraphics[width=.45\textwidth]{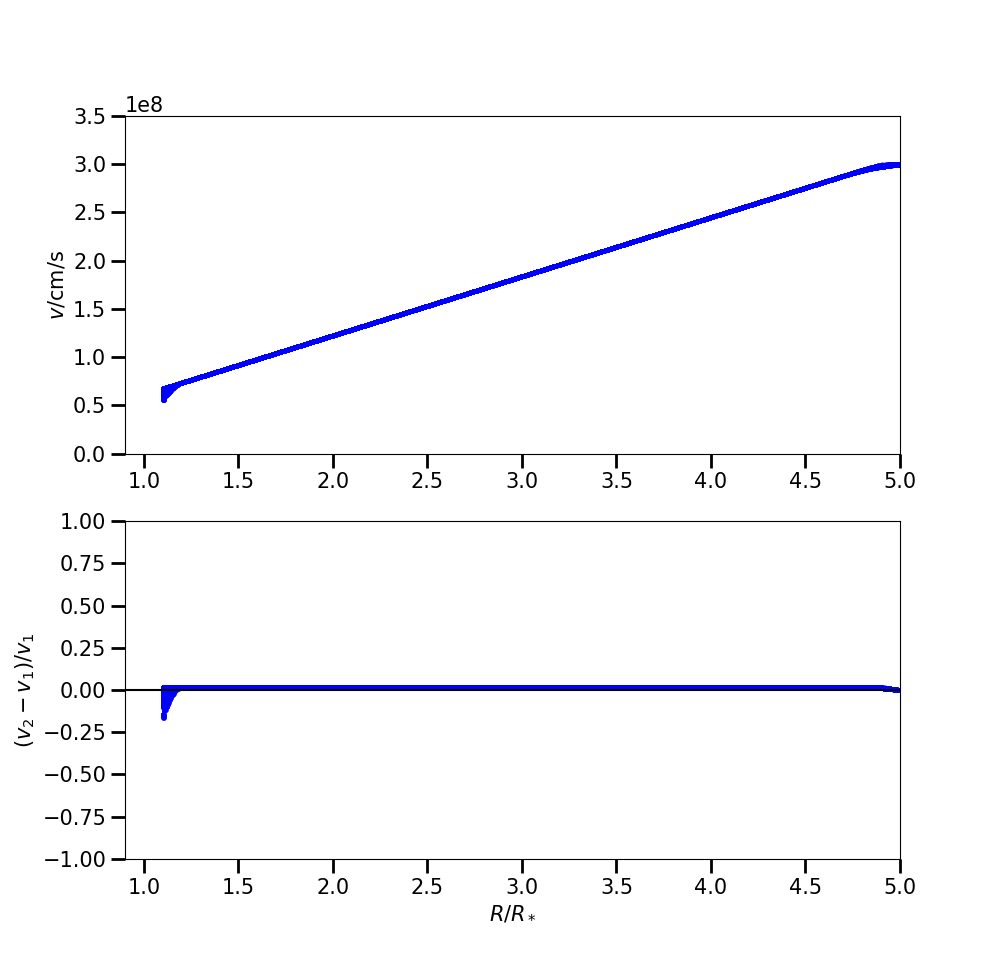}
 \includegraphics[width=.45\textwidth]{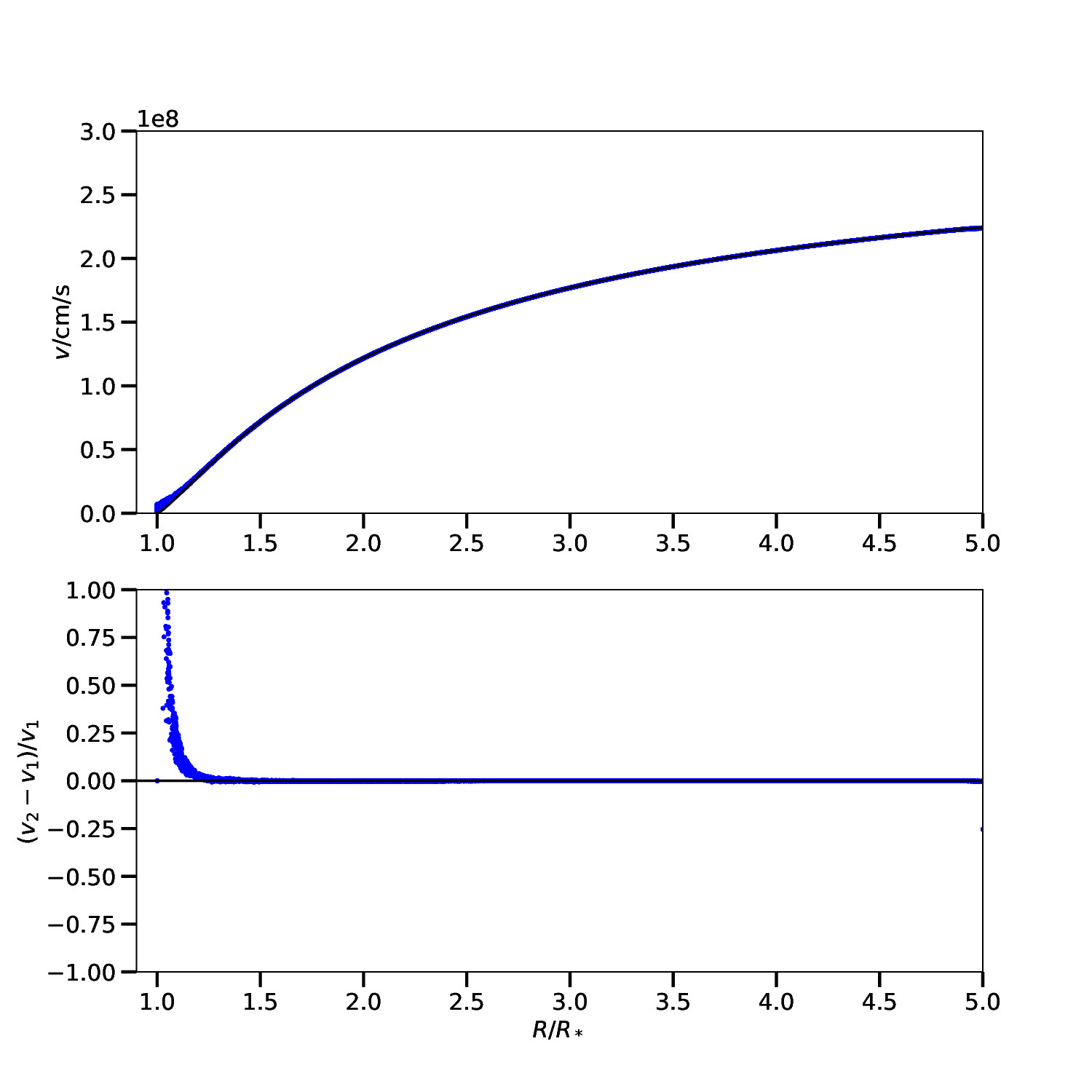}
 \caption{ A comparison of the numerical and analytical treatments of velocity,
  to assess the accuracy of the calculated velocity as a function of radius. The
  left panels are calculated using the homologous approximation
  \eqref{Eq:velhomol} and the right panels are calculated using the
  $\beta$-velocity law \eqref{Eq:beta3D}. The upper panels show the absolute
  magnitude of the velocity, and the lower panels show the relative difference
  in velocity magnitudes between the interpolated and analytical values.}
 \label{Fig:vel_beta}
\end{figure*}

The second test was performed using the velocity field defined by the
$\beta$-law; this law is expressed as
\begin{equation}
 \vel(r) = V_\infty\left(1-\frac{R_*}{r}\right)^\beta,
 \label{Eq:beta1D}
\end{equation}
where $\beta > 0$ is a parameter determining how quickly the wind is
accelerated. The velocity field has been sampled in discrete points, see the
Eq.~\eqref{Eq:hom_I}; $\vel_I$ is clearly equal to
\begin{equation}
 \vel_I = V_\infty\left(1-\frac{R_*}{\sqrt{x_I^2+y_I^2+z_I^2}}\right)^\beta,
 \label{Eq:beta3D}
\end{equation}
in the {\modgrid} cell with an index $I$. The calculations were performed using
the value of $\beta$ listed in the Table\,\ref{Tab:testing_hmodels}. The
interpolated values and relative errors are shown in the upper right and lower
right panels of Fig.\,\ref{Fig:vel_beta}, respectively. The values are very
close to the analytical function, it appears that there is no problem in this
case.

\subsection{Optical depth in lines}
The optical depth in the Sobolev approximation for an arbitrary velocity field is
calculated with the Eq.~\eqref{Eq:tauline}. However,
for velocity fields described in Section \ref{Sec:velfield} we can use the
simpler expression
\citepalias[][Eq.~(10)]{fisak2023}.
Both expressions will be used to test the accuracy of the numerical expression.

We combine equations (10) and (11) from \citetalias {fisak2023} and write the
equation for an optical depth in spherical symmetry in the form
\begin{equation}
 \tauline = \frac{c}{\nu_{lu}}\mathcal{K}f_{lu} n_l
  \left(1-\frac{n_u}{n_l}\frac{g_l}{g_u}\right)
  \left[\mu^2\frac{\text{d}\vel}{\text{d}r} + \left(1 - \mu^2\right) \frac{\vel}{r}\right]^{-1},
 \label{Eq:tauluss}
\end{equation}
here, $\mathcal{K}=\pi e^2/(m_\text{e}c)$.
In comparison to this equation,
Eq.\,\eqref{Eq:tauline} combined with \eqref{Eq:dsdnu} is
\begin{equation} 
 \tauline = \mathcal{K} f_{\mathrm{lu}}n_{l}
  \left(1 - \frac{n_u g_l}{n_lg_u}\right)
   \frac{s_+-s_-}{\nucmf_+-\nucmf_-}.
 \label{Eq:tauline2}
\end{equation}
For a given line, the equations \eqref{Eq:tauluss} and \eqref{Eq:tauline2}
should give the same value of $\tauline$ if the numerical evaluation of
\eqref{Eq:tauline2} is implemented correctly. Accuracy tests were performed on
velocity fields calculated using numerical interpolation. 20\,000 photon
packets were used for evaluation. When a photon packet encountered a line
during its passage through the atmosphere, the line's optical depths was
calculated using expressions \eqref{Eq:tauluss} and \eqref{Eq:tauline2}. We
refer to these as the `analytical' and `numerical' expressions, respectively.
These packets meet some line $N_\mathrm{points}$-times. We calculated a ratio
\begin{equation}
 R_\tau =
\frac{1}{N_\mathrm{points}} \sum_{i=1}^{N_\mathrm{points}}
 \frac{\tau_{i, \mathrm{line}}^{\mathrm{analytical}}}{\tau_{i,
 \mathrm{line}}^{\mathrm{numerical}}},
\end{equation}
which describes the accuracy of numerical calculation of the line optical
depth. In the ideal case, $R_\tau=1$. We test the values of $R_\tau$ for two
{\jednad} velocity fields: a homologous law and a $\beta$ velocity law.

\subsubsection{Homologous approximation}
For the case of a homologous approximation, the optical depth equation
\eqref{Eq:tauluss} simplifies into the form
\begin{equation}
 \tauline = \frac{c}{\nu_{lu}}\mathcal{K}f_{lu} n_l
 \left(1-\frac{n_u}{n_l}\frac{g_l}{g_u}\right)
 \frac{R_\mathrm{inf}}{V_\mathrm{inf}}.
 \label{Eq:taulspher}
\end{equation}
We calculated both values of optical depth
using \eqref{Eq:tauline2} (`numerical') and \eqref{Eq:taulspher} (`analytical'),
the values are equal in the ideal case $\tau_{i,
\mathrm{line}}^{\mathrm{numerical}}=\tau_{i,
\mathrm{line}}^{\mathrm{analytical}}$, however they slightly differ because of
numerical uncertainities.
In our calculation we got
$R_\tau = 0.959$ for $N_\mathrm{points}=2932$, which is a reasonable value.
%
\subsubsection{$\beta$-law}
Optical depth in the case of the $\beta$-velocity law is simply derivable from
the Eq.~\eqref{Eq:tauluss}, the optical depth is as follows
\begin{equation}
 \tauline = \frac{c}{\nu_{lu}}\mathcal{K}f_{lu} n_l
 \left(1-\frac{n_u}{n_l}\frac{g_l}{g_u}\right)
 \frac{r}{\vel(r)}
 \frac{1}{\left[\mu^2\left(\frac{R_*\beta}{r-R_*}-1\right)+1\right]}.
 \label{Eq:taulbeta}
\end{equation}
Here, $\mu$ is the cosine of the angle between the velocity vector and the
direction of the packet, and $\vel(r)$ is calculated using
Eq.~\eqref{Eq:beta1D}. For the $\beta$-velocity law we obtained $R_\tau =
0.997$ for $N_\mathrm{points}=64119$, which is again a reasonable value of
$R_\tau$.

\subsection{Calculated spectra}
In this section, we compare the emergent spectra calculated from our
simulations. Since spectral synthesis is our main goal in the developement of
new numerical models, it is essential that we verify the accuracy of spectra
calculated from models with numerically interpolated velocity fields. Our focus
will be on the previously discussed homologous approximation and the
$\beta$-law. Model parameters are listed in Table \ref{Tab:testing_hmodels} and
are the same for both velocity fields.
\subsubsection{Homologous approximation}
\label{Ssec:spec_homol}
\begin{figure*}
 \centering
 \includegraphics[width=.47\textwidth]{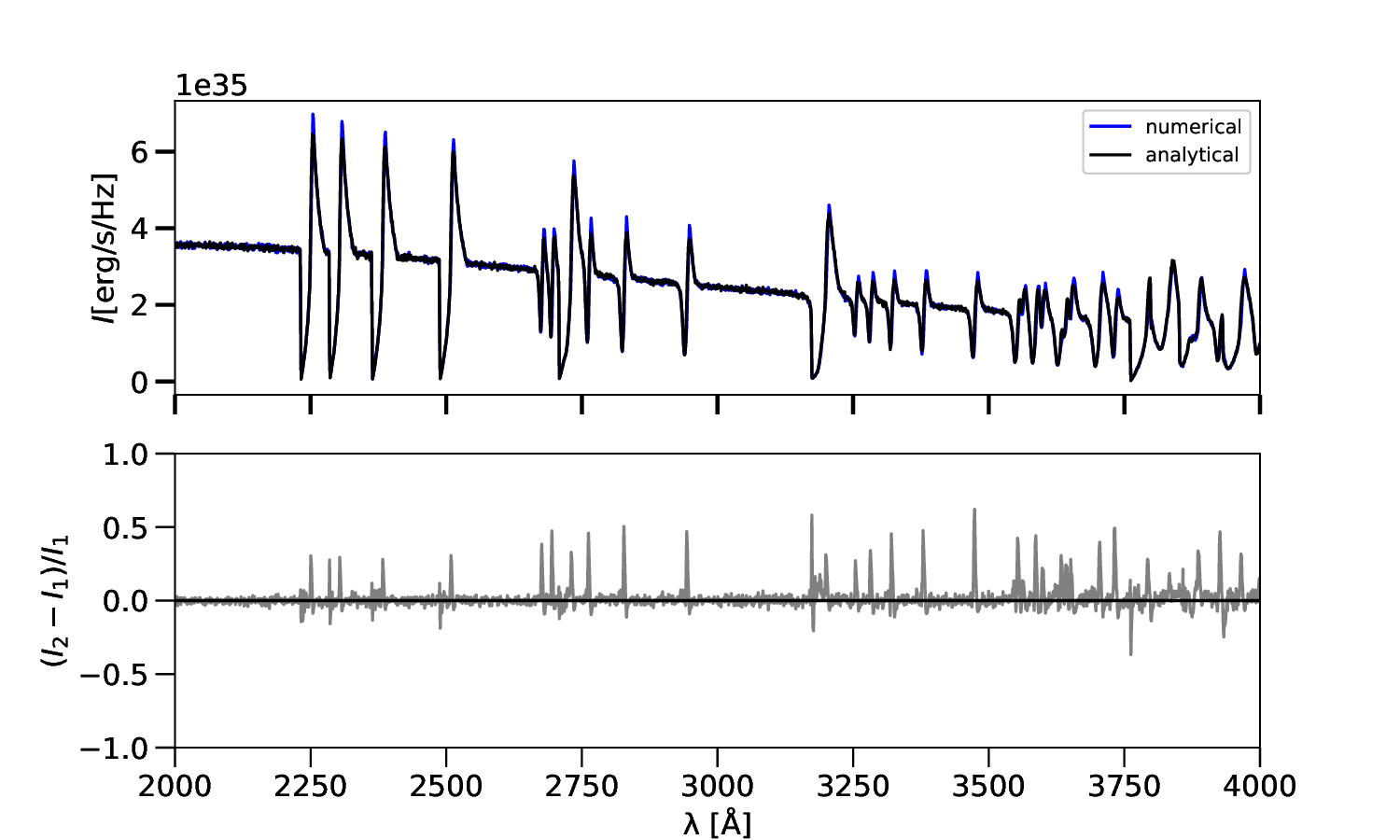}
 \includegraphics[width=.47\textwidth]{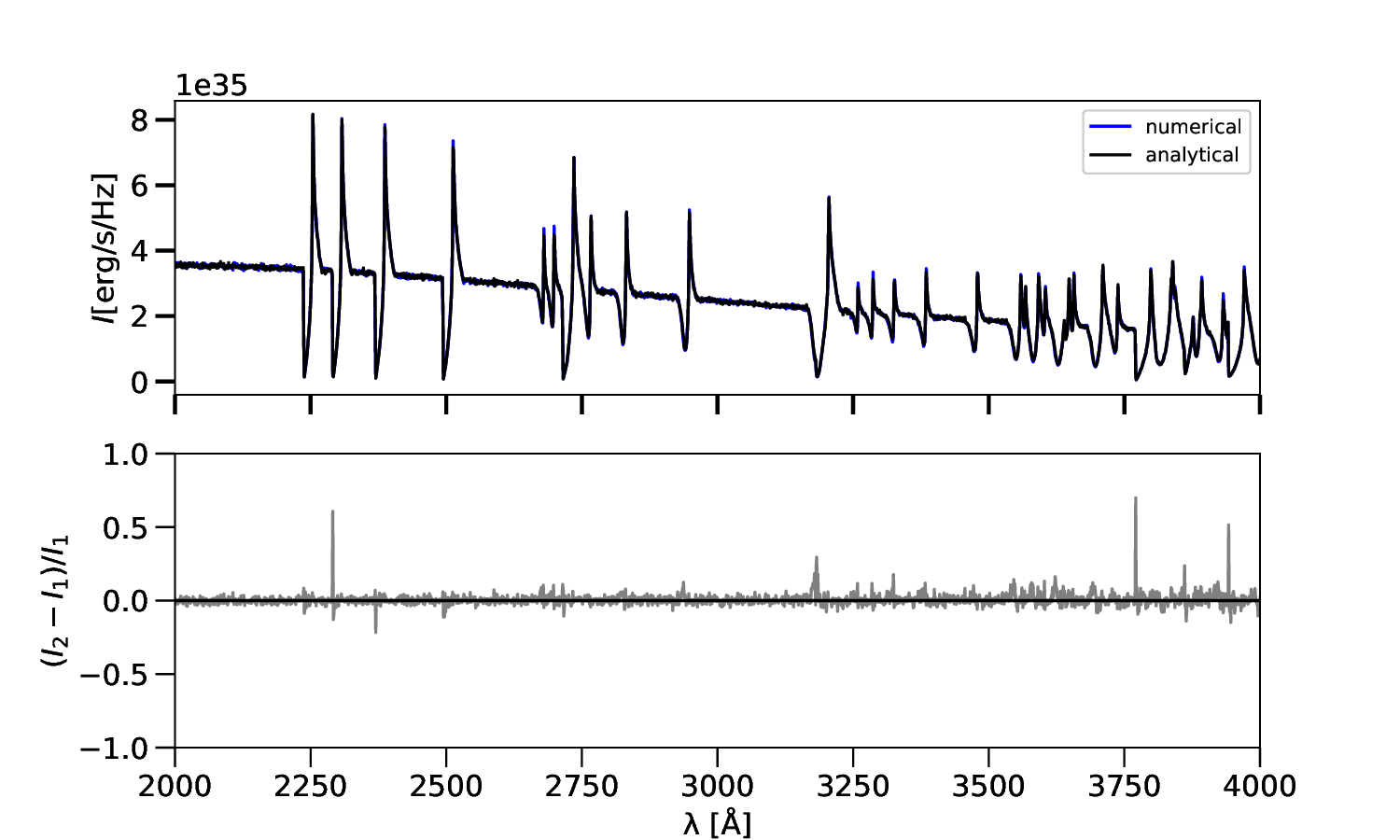}
 \caption{A comparison of the calculated spectra for two models. In the first model
  the velocity (and line optical depth) was interpolated from the {\trid}
  velocity points. In the second model the velocity (and line optical depth) was
  calculated using an analytical formula. \emph{Left panels}: homologous
  approximation, \emph{Right panels}: $\beta$ velocity law} 
  \label{Fig:spectratest}
\end{figure*}

We applied the equation of the homologous velocity field \eqref{Eq:velhomol}
and the corresponding optical depths \eqref{Eq:taulspher} or
\eqref{Eq:tauline2} for the analytical and numerical case, respectively. The
spectra and their relative differences for a homologous approximation are
plotted in the left panels of Fig.~\ref{Fig:spectratest}. The continua are in a
good agreement. The largest difference is caused by noise. There are many
spectral lines in the spectrum and there is no significant disagreement between
them. Noise decreases with increasing wavelength and the largest differences
occur in the emission profiles of the lines. Therefore the velocity
interpolation is calculated correctly in the case of the homologous
approximation.

\subsubsection{$\beta$-law}
The spectra for the $\beta$-velocity law were calculated using both analytical
\eqref{Eq:beta1D} and numerical velocity fields, as well as the line optical
depths using equations \eqref{Eq:taulbeta} and \eqref{Eq:tauline2} for the
analytical and numerical cases, respectively. The spectra are plotted in the
right panels of Fig.~\ref{Fig:spectratest}.

The spectra agree over a wide range of wavelengths, similarly to the case of
the homologous approximation, the main peaks lie in the emission parts of the
lines and the continuum noise is low.

\section{Radiative transfer solution for a {\dvad} hydrodynamic model} \label{Sec:applications}

To complement the accuracy analysis presented in the previous sections, we
demonstrate the calculation capabilities of our code by presenting a solution
to the problem of radiative transfer for a structure generated by a {\dvad}
hydrodynamic code.

\subsection{Brief description of the radiation-hydrodynamics code \mpiamrvac}

{\mpiamrvac} \citep[]{porthEtAl2014, xiaEtAl2018, keppensEtAl2020,
keppensEtAl2023} is a general-purpose multi-dimensional PDE-solver which has
specialised itself in solving the time-dependent equations of
(magneto-)hydrodynamics. \citep{moens2022b} developed a module to perform
radiation-hydrodynamic computations in {\mpiamrvac}, making use of the
flux-limited diffusion (FLD) formalism. This formalism requires the solution of
the 0th order, frequency-integrated radiative transfer equation, where FLD
offers an analytic closure relation between the radiative co-moving frame flux
and the radiation energy density.

The addition of the radiation subsystem to the hydrodynamic equations allows
for a more accurate treatment of both the source-terms in the momentum equation
(radiative forces), and radiative source-terms in the energy equation (heating
and cooling). Especially in a regime where the radiation field changes
dynamically as an effect of a complex density structure.

The code solves the RHD equations using a finite volume approach for the
advection terms, and a geometric multigrid solver \citep[][]{teunissen2019} for
the diffusive terms which are relevant for the FLD-closure. The finite volume
solver works on an oct/quadtree grid with adaptive mesh refinement. Multiple
Riemann solvers and timesteppers are available and described in
\citet{keppensEtAl2023}.

\citet{moens2022b} used the above described code to calculate a first set of
multi-dimensional RHD models of WR atmospheres. For this, the RHD equations
where solved in a local, box-in-wind Cartesian computational domain, where
spherical corrections where added for the radial $1/r^2$ factor in the
divergence terms. 

Later, \citet{debnath2024} used the same setup to study the turbulent
sub-photospheric regions in Ostars, \citet{udDoulaEtal2025} and
\citet{narechaniaEtal2025} continued this work by adding the effects of
magnetic fields.

Important in the setup of these atmosphere and wind models is the calculation
of the opacities. In the models used here, opacities are calculated as the sum
of the Rosseland mean opacities, which are important in the deep atmosphere,
and velocity-gradient enhanced opacities in the wind, making use of the Sobolev
approximation. This so-called hybrid opacity scheme has been described by
\citet{poniatowski2021, poniatowski2022}.

\subsection{Implementation to the Andy Antares code}\label{Ssec:implem2D}
In this subsection, we describe how the input model was implemented in the
code. When we implement a new model into the code, we have to test the
following steps
\begin{itemize}
 \item setting up the computational domain
 \item loading the input data
 \item transforming into Cartesian coordinates
 \item creating the {\propgrid} (test whether it describes the input model correctly)
\end{itemize}
\subsubsection{Setting the computational domain, loading the input data}

The input hydrodynamic model is two-dimensional, each point is defined by its
radius, $r$, and its lateral coordinate, $\vartheta$. This model describes a
{\dvad} part of the star, a `box in a wind'. The $r$ coordinate describes the
distance from the stellar centre. The model provides the mass density $\rho$,
temperature $T$, and two components of the velocity field $(v_r, v_\vartheta)$
in the $r$ and $\vartheta$ directions, respectively. To obtain a {\trid}
computational domain for radiative transfer calculations, we extend this `box
in a wind' model to describe the spherical star. First, we convert the {\dvad}
rectangular model into an annular sector of lateral width $\vartheta_{\max} =
\pi/n$, where $n$ is a natural number. Then, we place $n$ copies of this sector
periodically side by side to form an annular sector of lateral length $\pi$
within a plane intersecting the centre of the star. Finally, we rotate this
annular sector around the $z$-axis to obtain the entire stellar envelope (a
shell). For our calculations, we chose $n = 4$.

\subsubsection{Connection of {\propgrid} with the model points}
\begin{figure*}
 \begin{center}
 \includegraphics[width=.45\textwidth]{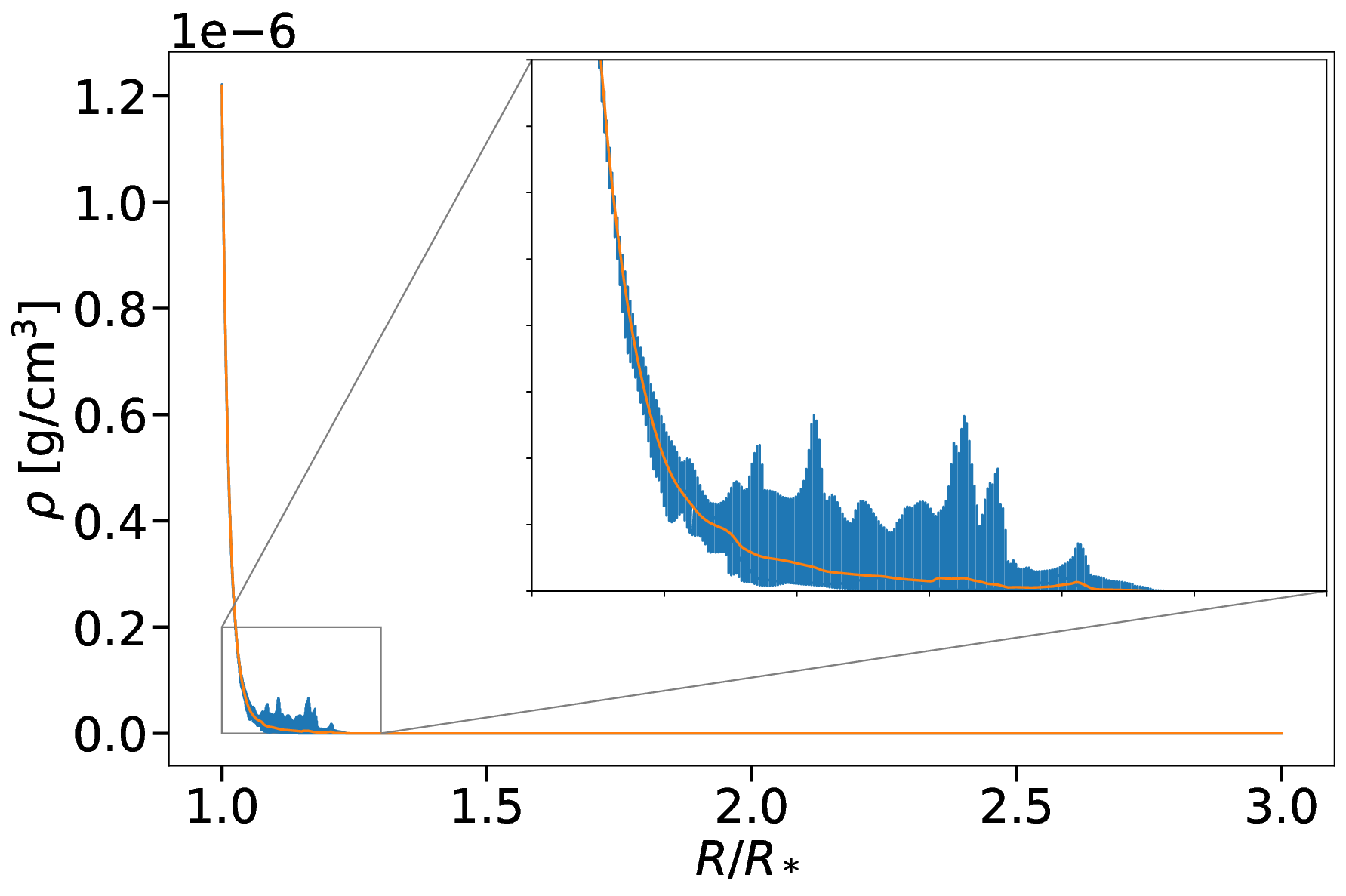}
 \includegraphics[width=.45\textwidth]{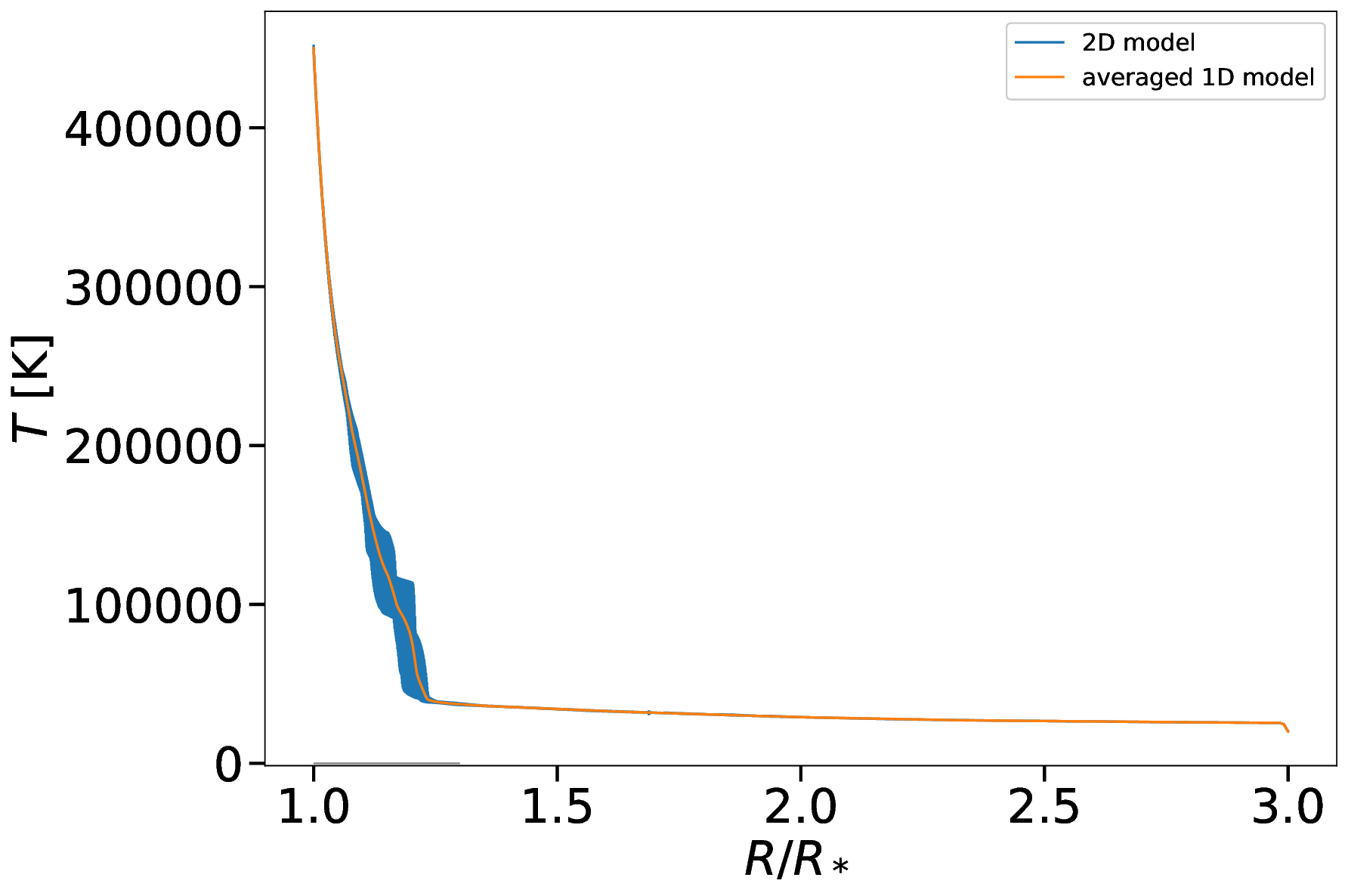}
 \includegraphics[width=.45\textwidth]{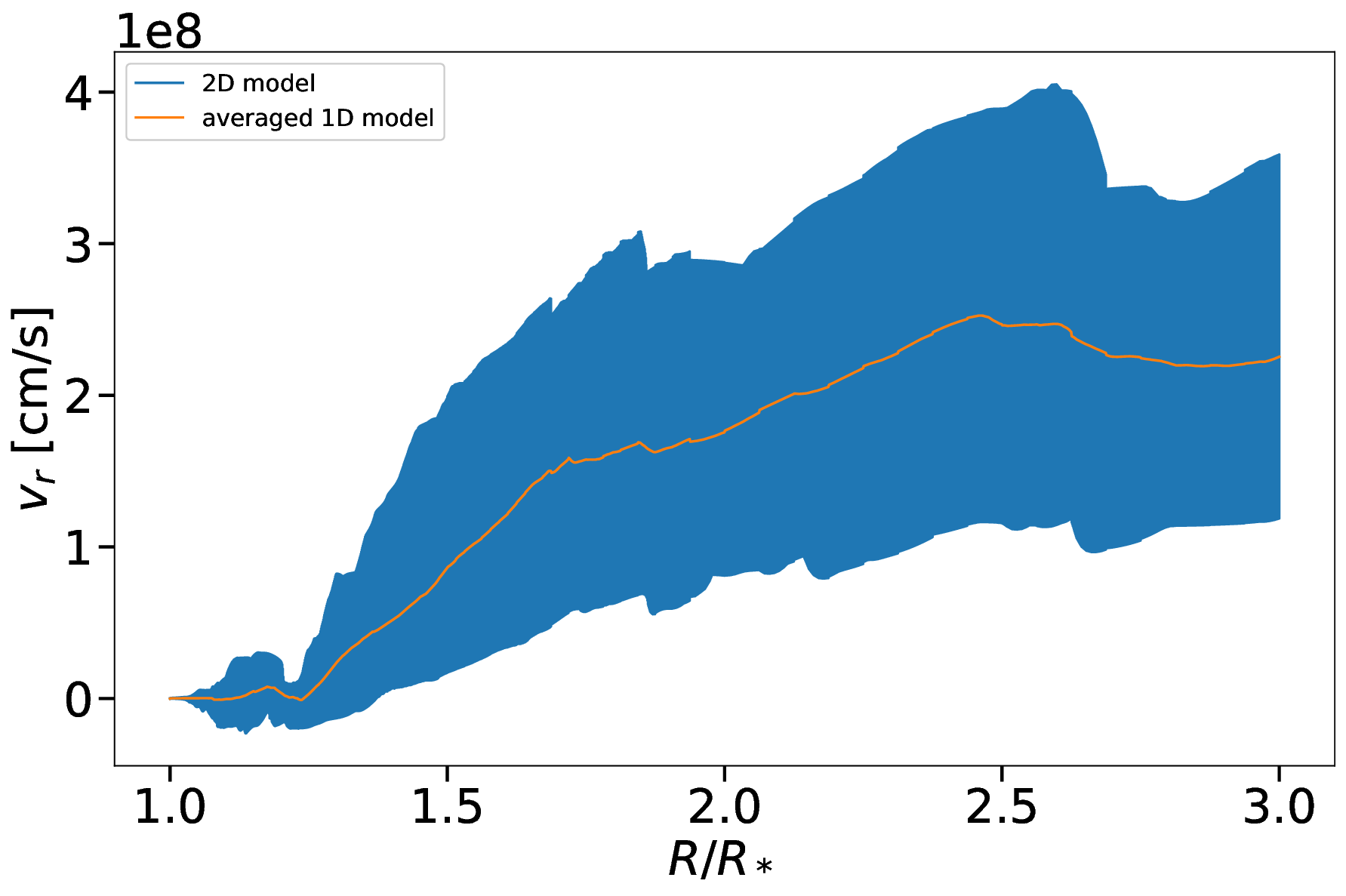}
 \includegraphics[width=.45\textwidth]{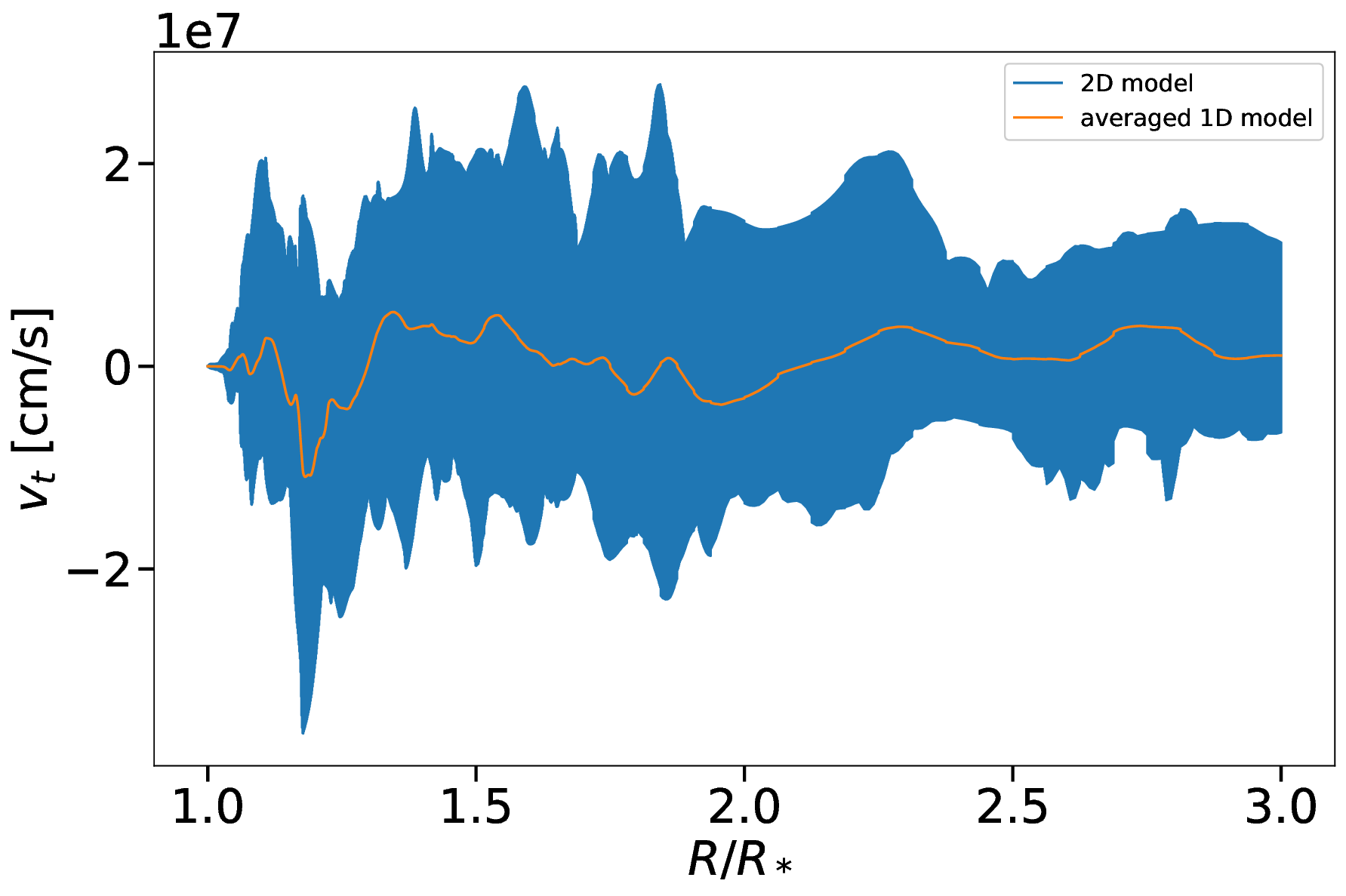}
 \end{center}
 \caption{The input model (density, temperature, and the radial and lateral
 velocities) calculated using the {\mpiamrvac} hydrodynamic code.}
 \label{Fig:input2dmodel}
\end{figure*}
The velocity vector of the {\modgrid} cell $I$ is defined by its radial $(v_r)_I$
and latitudinal $(v_\vartheta)_I$ components, which must be evaluated for each
{\propgrid} cell centre $J$ with position $r_J$ and $\vartheta_J$ and $\varphi_J$
and the Cartesian components of this vectors are
\begin{equation}
 \begin{pmatrix}
  v_x \\
  v_y \\
  v_z
 \end{pmatrix}_J
 =
 \begin{pmatrix}
  (\vel_r)_I\cdot \sin(\vartheta_J) \cdot \cos(\varphi_J) +
   (\vel_{\vartheta})_I\cdot\cos(\vartheta_J)\cdot\cos(\varphi_J)\\
  (\vel_r)_I\cdot \sin(\vartheta_J) \cdot \sin(\varphi_J) +
   (\vel_{\vartheta})_I\cdot\cos(\vartheta_J)\cdot\sin(\varphi_J)\\
  (\vel_r)_I\cdot \cos(\vartheta_J) + (\vel_{\vartheta})_I\cdot\sin(\vartheta_J)
 \end{pmatrix}.
\end{equation}

The {\mpiamrvac} model implemented was calculated in {\dvad} geometry, which
brings a different treatment from the {\jednad} and {\trid} cases. The model
points are defined by radius ($r$) and lateral coordinate ($\vartheta$). We can
get all the full information we need to calculate spectra. We load model points
and for each point we get: radius, temperature, density, angular and lateral
velocities. The next step is to set up the computational domain: the lower and
the upper limits which are defined by the points with the smallest radius and
largest radius. The {\propgrid} is created, and each {\propgrid} cell is
assigned to a {\modgrid} cell. Since we know the cartesian coordinates, we
calculate the lateral coordinates of each {\propgrid} cell with a position
$\vec{r}_J=(x_J, y_J, z_J)$ as $$r_J=\sqrt{x^2_J+y^2_J+z^2_J},$$ and lateral
coordinates $$\vartheta_J = \frac{z_J}{\sqrt{x^2_J+y^2_J+z^2_J}},$$ we can
calculate the distance from a {\modgrid} cell with coordinates $r_I$ and
$\vartheta_i$ as
\begin{equation}
 \delta = \sqrt{r^2_J+r^2_I - 2r_J\cdot r_I \cdot \cos(\vartheta_J -
 \vartheta_I)},
\end{equation}
we are seeking the {\modgrid} cell with the shortest distance from the
{\propgrid} centre. We can run several tests to see if the {\propgrid} we
created accurately describes the {\modgrid}. One important parameter is a
relative coverage described in \citetalias[Eq.~(41)]{fisak2023}, which
describes a ratio of the number of {\modgrid} cells that are connected to at
least one {\propgrid} cell to the total number of {\modgrid} cells. 


\subsection{Preliminary results of our simulations}

The Tab.~\ref{Tab:2D_param} contains important {\modgrid} and {\propgrid}
parameters including the lower boundary conditions.
%
\begin{table*}
 \centering
 \begin{tabular}{p{5cm} c c}
  \hline
   \textbf{parameter} & \textbf{{\dvad}} & \textbf{av. {\jednad}}\\
  \hline
   lower boundary $R_*/R\odot$ & 160  & 160\\
   upper boundary $R_\infty/R\odot$ & 480  & 480\\
   $\vartheta_\mathrm{max}$ & $\pi/4$ & -- \\
  \hline
   number of {\modgrid} cells & 1048576  & 2048\\
   number of regular {\propgrid} cells & 700 $\times$ 700 $\times$ 700 &
     700 $\times$ 700 $\times$ 700 \\
   number of {\propgrid} cells & 343 000 000 & 343 000 000 \\
   relative covering &  40 \% & 100 \% \\
  \hline
   number of packets & 15 000 000 & 15 000 000 \\
   effective temperature [K] & \multicolumn{2}{c}{30 000 or variable} \\
  \hline
 \end{tabular}
 \caption{Numerical parameters of the calculated model.}
 \label{Tab:2D_param}
\end{table*}

\begin{figure*}
 \centering
 \includegraphics[width=.9\textwidth]{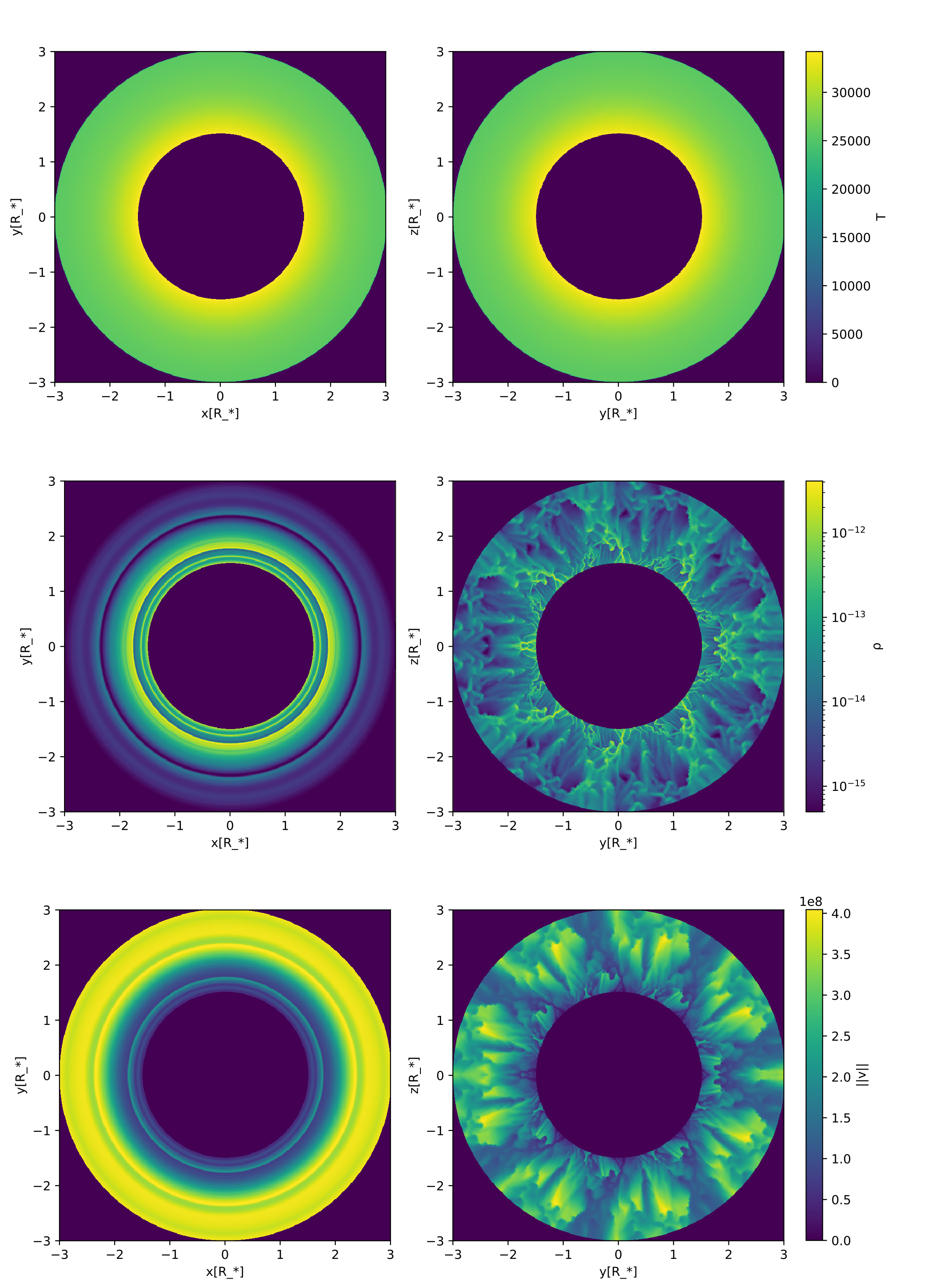}
 \caption{Visualisation of the entire {\propgrid} cut. A total of $700 \times
  700$ square cells are plotted in each figure. Each square represents one
  {\propgrid} cell. The physical quantities are constant within the {\propgrid}
  cell (except for the velocity field). \textbf{Columns}: \emph{Left}: $xy$
  plane, the axis of symmetry is perpendicular to the plane of the plot, as can
  be seen from the plot, \emph{right}: \emph{xz} plane, the axis is parallel to
  the plane of the plot. 
  \textbf{Rows}: 1. temperature, 2. density, 3. magnitude of velocity.}
 \label{Fig:propmod1}
\end{figure*}

The distribution of density in the calculated model is depicted in
Fig.~\ref{Fig:propmod1}. The grid is fine enough to resolute the included
inhomogeneities. 

\begin{figure*}
 \centering
 \includegraphics[width=.75\textwidth]{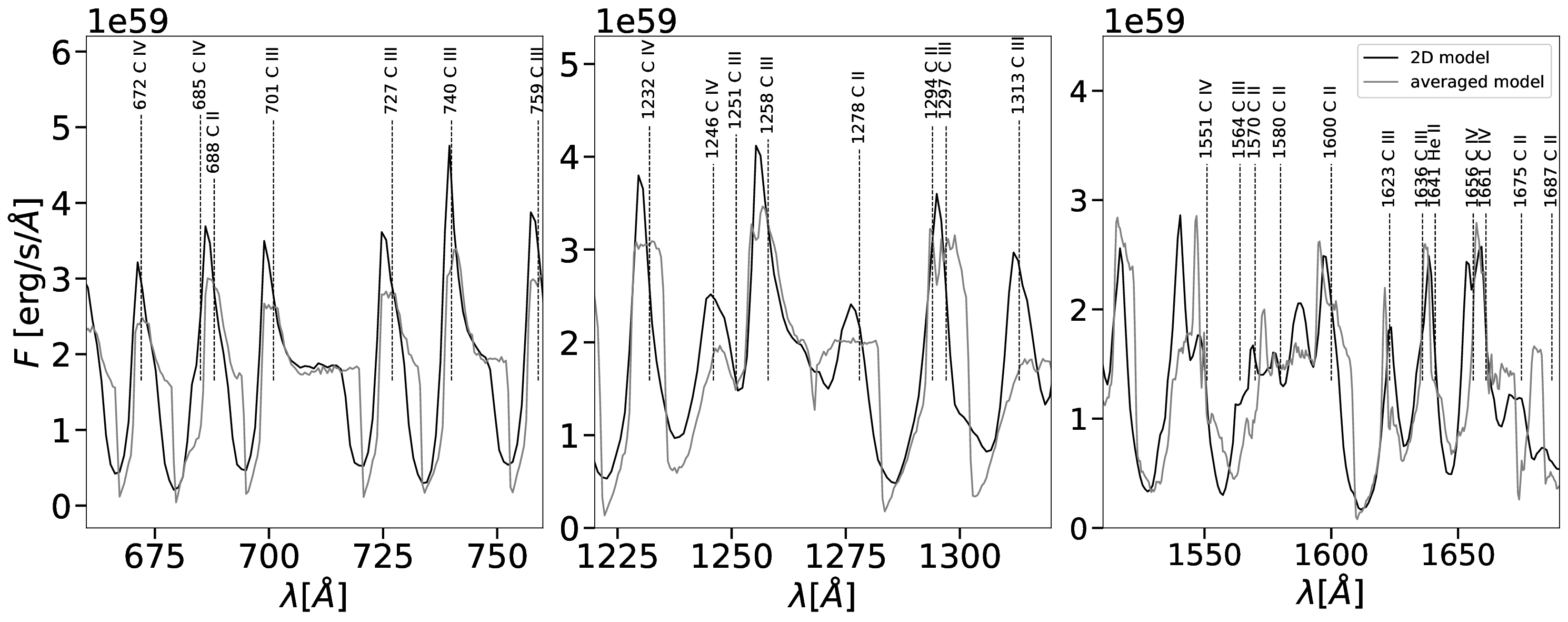}\\
 \includegraphics[width=.75\textwidth]{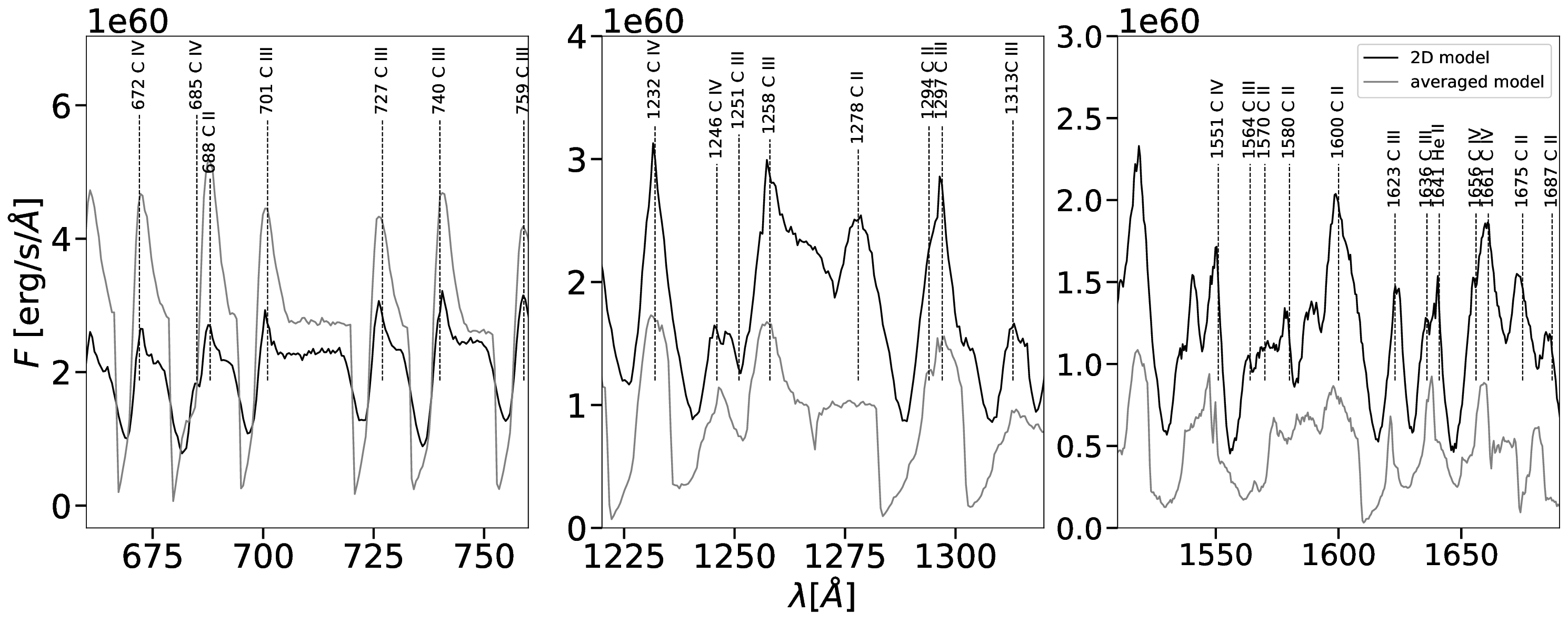}
 \caption{Calculated spectra of two models from {\mpiamrvac}: {\dvad} and
 the same model averaged in $\theta$. The difference between upper and lower
 spectrum is the lower boundary effective temperature. \emph{Above}:
 $T_\mathrm{eff} = 30000\,\mathrm{K}$, \emph{below}: effective temperature is
 defined by the temperature of the model in the initial packet point.}
 \label{Fig:spec_hydromodel2}
\end{figure*}

We calculated the very first testing spectra of a homogeneous wind consisting
of Hydrogen (73.46 \%), Helium (24.85 \%) and Carbon (0.29 \%) by mass. To
evaluate the importance of including of inhomogeneities, we created a {\jednad}
model by averaging the physical quantities from the {\dvad} model in the
lateral coordinates at each radial point, and solved radiation transfer for
this {\jednad} model. Using the same geometric dimensions and the lower
boundary conditions we calculated the spectrum emerging from the {\jednad}
model and compared it with the {\dvad} model spectrum in
Fig.~\ref{Fig:spec_hydromodel2}. Naturally, the two spectra differ. The
{\jednad} model, which is an average of the {\dvad} model, smooths the
inhomogeneities in the medium, as it is depicted by the orange lines in
Fig.\,\ref{Fig:input2dmodel}. Consequently, it produces an emergent spectrum
that differs from that emerging from the more exact {\dvad} input model. This
highlights the importance of accurately resolving atmospheric inhomogeneities
and including them in models. A similar conclusion was found for a simpler
model by \cite{TichyKubat2019}.

The figures (upper and lower) in the Fig.~\ref{Fig:spec_hydromodel2} differ in
the lower boundary conditions. The upper spectrum was calculated for an
assumption of Plank body radiation in the lower boundary ($r=R_*$). The spectra
below were calculated also for Plank distribution function in the lower
boundary, but the effective temperature has been defined from the {\modgrid},
hence the the distribution was depending also on the initial position of the
packet.

\section{Conclusions}
We present a Monte Carlo radiative transfer code for the spectral synthesis.
Our code is designed to calculate radiative transfer in multidimensional input
models ({\dvad} and {\trid}) including general velocity fields (non-monotonic).

We have described the used numerical methods. The interaction in lines is
simply generalised for non-monotonous velocity fields. In addition to the CMF
redshift of a packet, the CMF blueshift is also included. The velocity field is
linearly interpolated from the discrete points defined by the model. We
performed interpolation tests using two simple velocity profiles: the
homologous approximation (linear) and the $\beta$-law velocity profile. For
each case, the interpolated velocity profile was compared with its analytical
value. The differences become larger near the upper ($r\approx R_*$) and the
lower boundary ($r\approx R_\infty$) limits. The middle parts of the velocity
profiles are calculated with an error lower than 5\,\%. We compared the
emergent spectra of all the velocity profiles. The spectra differ mainly in the
emission parts of the lines, especially in the case of the homologous
approximation. The continua fit well. 

We adapted our code to compute the spectrum emerging from the {\dvad}
hydrodynamic model {\mpiamrvac}, which is arranged in the form of a ``box in
the wind''. The implementation of the model in our code is described in
Section~\ref{Ssec:implem2D}. The main challenge lies in creating a {\propgrid}
that encompasses all the inhomogeneities, which requires the inclusion of each
{\modgrid} point in the {\propgrid}. For our simulations, we created a regular
{\propgrid} and calculated the spectra. The spectrum of the {\dvad} model is
compared with the spectrum of the {\jednad} model derived from the {\dvad}
model by averaging quantities in the lateral coordinates. The two spectra
differ significantly in terms of their line details, which highlights the
importance of solving the problem using a multidimensional approach.

The {\dvad} model also introduces a new effect to packet propagation: in the
case of monotonic velocity fields, red-shifting of the packet was the only
possibility. In more complex velocity fields, blue-shifting of the packet is
also possible. The effect of velocity fields with
$\frac{\text{d}||v||}{\text{d}r} < 0$ is usually neglected in the radiative
transfer codes. However, it was straightforward to incorporate this effect into
our code.

Radiative transfer is solved under several assumptions. The first limitation is
the {\propgrid} used. Although our code supports adaptive mesh refinement, we
only used a regular grid here which reduced the number of used {\modgrid}
points used by the model to 40\,\%. However, it was still accurate enough to
include inhomogeneities and clumping, as shown in Fig.~\ref{Fig:propmod1}. We
compared the spectra calculated in the {\dvad} model and {\jednad} models with
the averaged quantities from the {\dvad} model. The calculated spectra differ,
mainly in the second case (the lower spectrum of
Fig.~\ref{Fig:spec_hydromodel2}). Therefore it is important to calculate
{\dvad} and {\trid} models. The differences would be more significant if
spectra were calculated from a specific perspective, which has not yet been
done.

\begin{acknowledgements}

JF was supported by the PPLZ programme of the Czech Academy of Sciences.
Computational resources were supplied by the project "e-Infrastruktura CZ"
(e-INFRA CZ LM2018140 ) supported by the Ministry of Education, Youth and
Sports of the Czech Republic. JK and BK acknowledge support from the Grant Agency of the Czech Republic (GA\v CR 25-15910S). The Astronomical Institute of the Czech Academy of Sciences in Ondřejov is supported by the
project RVO: 67985815.
\\ \emph{Special thanks from JF:}
I would also like to thank a great teacher from my secondary school, Lum\'{i}r
Svoboda. He was a really brilliant teacher who motivated me to study maths and
physics more intensively. I always recall those times at Elgartova School
fondly.

\end{acknowledgements}
\bibliographystyle{aa}
\bibliography{./Literatura}
\appendix
\section{The trilinear interpolation}
\label{Ap:trilinintp}
\begin{figure}
 \centering
 \includegraphics[width=.45\textwidth]{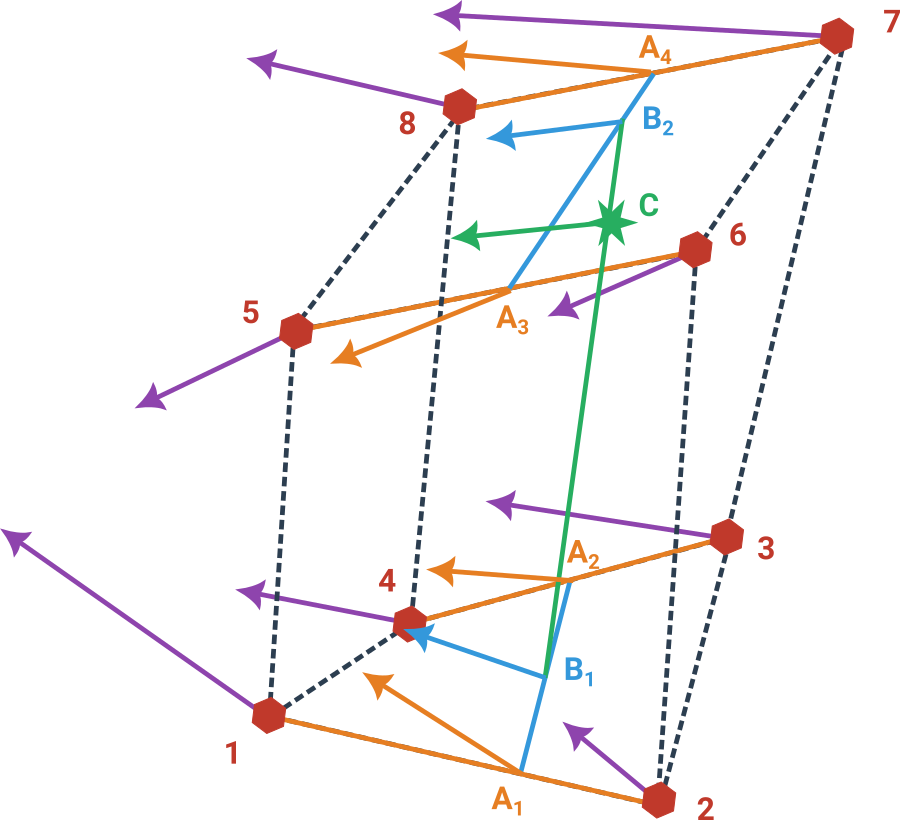}
 \caption{Scheme of the trilinear interpolation. We aim to determine the value
 at the green point (C). The interpolated physical quantity is defined in the eight
 red points placed in the red hexagonals. Using linear interpolations, new
 values are calculated in four orange points (A$_1\cdots$ A$_4$), using these orange
 points we obtain values in the blue points (B$_1$, B$_2$), and, finally, we
 calculate the value in the green point C.}
 \label{Fig:trilinear}
\end{figure}

The trilinear interpolation consists of linear interpolations performed
successively in three dimensions, it is described in \cite{numrecipes} and
depicted in Fig.~\ref{Fig:trilinear}.

A non-trivial part is a determination, which {\propgrid} cells are neighbouring
the current cell and can be used for this algorithm. This calculation is
described in the Section \ref{Ssub:intp_modgrid} and depicted in the
Fig.~\ref{Fig:interpvel}. In the beginning, an
octant of the {\propgrid} cell where the value in the green point from
Fig.~\ref{Fig:trilinear} is determined. 

Let us assume two points: A$_1$ and A$_2$ at positions $\mathbf{r}_1$ and
$\mathbf{r}_2$ containing velocity vectors $\velb_1$ and $\velb_2$,
respectively. We want to find a velocity vector at a specific point on a
straight line between these two points $\mathrm{A}_1$ and $\mathrm{A}_2$. The
equation of the straight line is equal to
\begin{equation}
 \mathbf{r}(\delta) = \mathbf{r}_1 + \frac{\mathbf{r}_2 -
 \mathbf{r}_1}{|\mathbf{r}_2 - \mathbf{r}_1|}\cdot \delta.
  \label{Eq:strLine}
\end{equation}
Here, $\delta$ is the parameter of the straight line (distance from the point
$\mathbf{r}_1$). Assuming the following conditions $\mathbf{r}_1$, it is
$\delta = 0$, in the second point $\delta = |\mathbf{r}_2-\mathbf{r}_1| = d$,
we can write the interpolated velocity is in the form
\begin{equation}
 \velb(\delta) = \frac{\velb_1 - \velb_2}{d}\cdot \delta +
 \velb_1.
\end{equation}
The value of the parameter $\delta$ is obtained from the
Eq.~\eqref{Eq:strLine}, where we put a selected coordinate ($x$, $y$, or $z$)
calculated. The interpolation is independent of the order of this choice.

Neighbouring {\propgrid} cells are chosen according to a relative position of
the green point to the current {\propgrid} centre -- those {\propgrid} cells
have common bound, edge, or corner with the occupied octant of the current
{\propgrid} cell.

\end{document}